\documentclass[aps,prl,superscriptaddress,showpacs,twocolumn,floatfix]{revtex4}

\usepackage{bm}
\usepackage{graphicx}
\usepackage{epsfig}
\usepackage{color}

\begin{document}

\newcommand{\qav}[1]{\left\langle #1 \right\rangle}
\newcommand{\myT}{\Gamma}
\newcommand{\rem}[1]{}
\newcommand{\refe}[1]{(\ref{#1})}
\newcommand{\fige}[1]{Fig.~\ref{#1}}
\newcommand{\refE}[1]{Eq.~(\ref{#1})}
\newcommand{\beq}{\begin{equation}}
\newcommand{\eeq}{\end{equation}}
\newcommand{\beqa}{\begin{eqnarray}}
\newcommand{\eeqa}{\end{eqnarray}}
\newcommand{\cg}{\check g}
\newcommand{\inc}{{\rm inc}}
\newcommand{\Pc}{{\cal P}}
\newcommand{\larghezza}{7.8cm}
\newcommand{\cG}{\check{\cal G}}

\title{Frequency dispersion of photon-assisted shot noise in mesoscopic conductors}

\author{D. Bagrets}
\affiliation{Institut f\"ur Theoretische Festk\"orperphysik,
Universit\"at Karlsruhe, 76128 Karlsruhe, Germany}
\affiliation{Forschungszentrum Karlsruhe, Institut f\"ur Nanotechnologie,
76021 Karlsruhe, Germany}
\author{F. Pistolesi}
\affiliation{Laboratoire de Physique et Mod\'elisation des Milieux
    Condens\'es, CNRS-UJF B.P. 166, F-38042 Grenoble, France}

\date{\today}

\begin{abstract}
We calculate the low-frequency current noise for AC biased mesoscopic
chaotic cavities and diffusive wires.
Contrary to what happens for the admittance, the frequency dispersion
is not dominated by the electric response time (the "RC" time of the circuit),
but by the time that electrons need to diffuse through the structure
(dwell time or diffusion time).
Frequency dispersion of noise stems from fluctuations of the Fermi distribution
function that preserve charge neutrality.
Our predictions can be verified with present experimental technology.
\end{abstract}
\pacs{
72.70.+m, 
73.23.-b, 
73.50.Td, 
73.50.Pz. 
}

\maketitle

Charge neutrality in mesoscopic devices is enforced by Coulomb
interaction on a time scale ($\tau_{RC}$) given by the product of
the resistance and the capacitance scales.
The typical time ($\tau_D$) for non interacting electron motion is
instead fixed by either the dwell or diffusion time, depending on
the transparency of the interfaces between the system and the
electrodes.
The typical physically realized situation is $\tau_{RC}\ll \tau_D$:
electrons, which would normally slowly diffuse, are pushed to run by the
electric fields that they are generating themselves by piling up charge.
The consequence is that the typical response time of the
device is $\tau^{-1}=\tau_{RC}^{-1}+\tau_{D}^{-1}\approx \tau_{RC}^{-1}$.
This has been shown for the frequency dependence of both the
admittance \cite{buttiker:1993,naveh:1999} and the
noise \cite{nagaev:1998,golubev:2004,nagaev:2004,hekking:2006} in
mesoscopic chaotic cavities and diffusive wires.
The inverse of the diffusion time appears instead as the relevant energy
scale for the voltage or temperature dependence for both the conductance
and the noise in superconducting/normal metals hybrid
systems \cite{volkov:1993,hekking:1994,belzigWire:2001,houzet:2004}.
Indeed, in this case the energy dependence is due to
interference of electronic waves that does not induce charge
accumulation in the systems.
To our knowledge, in normal metallic structures a dispersion on the
inverse diffusion time scale has been predicted so far only for the
third moment of current fluctuations \cite{nagaev:2004} and for the
{finite frequency} thermal noise response to an oscillating
heating power \cite{reulet:2005}.
An alternative, and less investigated possibility, is to study the
{ low frequency} current noise as a function of the frequency
$\Omega$ of an external AC bias.
The noise for a quantum point contact was calculated a decade
ago \cite{lesovik:1994} and later measured \cite{schoelkopf:1997,reydellet:2003}.
Since the quantum point contact is very short, electron diffusion
does not introduce any additional time scale in the problem and the resulting
frequency dispersion is simply linear \cite{lesovik:1994}.
More recently, the noise for an AC biased chaotic cavity was
considered \cite{rychkov:2005,polianski:2005} in the limit of small
fields $eV/\hbar\Omega \ll 1$ ($V$ is the amplitude of the AC bias
and $e$ is the electron change).
The authors of Ref.~\cite{rychkov:2005} found that the noise
disperse only on the $\tau$ scale, that is the combination of
the diffusion and the electric response time.
This should be contrasted with the result of Ref.~\cite{altshuler:1994}
where the noise in a diffusive wire was
studied when the conductor was biased by a short voltage pulse.
Even if the dispersion of the noise is not considered in that paper,
the authors find that the ratio of the diffusion time to
the time duration of the pulse may affect the observed noise.
This indicate that a dispersion of the noise on the external frequency
could be present.
%

%
%
%
%
\begin{figure}
\begin{center}
\includegraphics[width=2.5in]{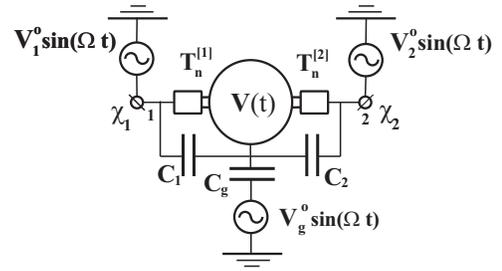}
\caption{Schematic of the system. A chaotic cavity
connected to the electrodes through arbitrary coherent connectors
characterized by a set of transparencies $\{ T_n \}$. }
\label{fig1}
\end{center}
\end{figure}

In this paper we derive analytical expressions for the photon-assisted
noise in chaotic cavities [\refE{finalS} in the following] and numerical results
for diffusive wires.
We show that the noise depends on the AC frequency $\Omega$ on the
diffusion time scale.
We take into account the Coulomb response that guarantees charge
neutrality over times longer than $\tau_{RC}$.
Within our calculation, the admittance has no structure at
$\Omega \approx  1/\tau_D$, while the derivative of the noise with
respect to the frequency shows a clear maximum.

We start by considering a chaotic cavity (see Fig. \ref{fig1}) connected to
two metallic leads through two barriers characterized by a set of
transparencies $\{ T_n^{k} \}$ where $k$ takes the value 1 or 2.
We define the conductances $G_k = G_Q \sum_n T_n^{k}$, the total conductance
$G=G_1G_2/(G_1+G_2)$, and a dwell time $\tau_D = 2\pi\hbar G_Q/(G_1+G_2) \delta$,
where $\delta$ is the level spacing of the cavity and $G_Q=e^2/(2\pi\hbar)$.
We assume that the dwell time is much shorter that the inelastic
time. We also assume that $G_\Sigma=G_1+G_2\gg G_Q$ so that we can use
semiclassical theory to describe transport and neglect Coulomb blockade effects.
The cavity is coupled capacitively to the leads through the
capacitances $C_k$ and to a gate through $C_g$.
One can then define the typical electric response time as
$\tau_{RC}=C_\Sigma/G$ where $C_\Sigma=C_1+C_2+C_g$.
If linear dimension of the cavity $L$ are much larger than the Fermi
wavelength $\lambda_F$, the ratio $\tau_{RC}/\tau_{D} \approx \delta/E_C \approx
(\lambda_F/L)^{d-1} \ll 1$ where $E_C=e^2/C_\Sigma$ is the Coulomb
energy and $d=2$ or $3$ is the dimension of the cavity.
Three different time dependent voltage biases are applied to the
gate and the two contacts ($V_g$, $V_1$, and $V_2$), clearly the
current depends only on two voltage differences, we keep the three
voltages to simplify the notation.
The cavity resistance is negligible with respect to the contact
resistances so that we can assume a uniform electric potential
inside.

We begin by considering a very simple model of charge transport,
where diffusion and electric drift are described classically.
The charge $Q$ in the cavity is related to the electric potential
$V$ of the cavity itself by the relation $Q = \sum_k C_k(V-V_k)$,
where $k=1$, $2$, and $g$.
The equation of motion for $Q$ is then
\beq
    \partial_t Q + Q/\tau_D + I_1+I_2=0,
    \label{diffusion}
\eeq
where $I_k = G_k(V-V_k)$.
The equation can be conveniently solved in terms of $V(\omega)$, the
Fourier transform of the cavity potential:
\[
    V(\omega) = {\tau/\tau_D \over 1 -i \omega \tau}
    \sum_{k=1,2,g}
    V_k(\omega)
    \left[
    \alpha_k {\tau_D \over \tau_{RC}}
    +
    (1- i \omega \tau_D)
    {C_k\over C_{\Sigma}}
    \right],
\]
where we defined $\alpha_k=G_k/G_\Sigma$ for $k=1$ or $2$, and
$\alpha_g=0$.
When $\tau_{RC}\ll \tau_D$,
the response time $\tau=\tau_{RC}\tau_D/(\tau_{RC}+\tau_D)$
is simply given by $\tau_{RC}$.
The admittance has a similar frequency dependence.
For $\tau_{RC} \ll \tau_{D}$ and $\omega \tau \ll 1$ one thus finds
the very simple result
\beq
    V(\omega) = \sum_k \alpha_k V_k(\omega),
    \label{voltage}
\eeq
 and
$I_1(\omega)=-I_2(\omega)=G_\Sigma \alpha_1 \alpha_2 (V-V_1)$:
charge neutrality is perfectly enforced for low frequency drive.
As it follows from \refE{diffusion}, only $I_k$ is finite,
since $Q$ and $\partial_t Q$ vanish in this limit.
In the following we will consider this experimentally relevant low
driving-frequency limit $\Omega \tau_{RC} \ll 1 $, which will enable
us to use the $\omega \ll 1/\tau_{RC}$ response for $V(\omega)$.
%


Let us now consider the noise.
We use the semiclassical description of non-equilibrium transport
provided by the time dependent Usadel equations \cite{usadel:1970,larkin:1986}:
\begin{equation}
    D \nabla (\cG \nabla \cG) +
    \left[ i\hbar  (\partial_{t_1} + \partial_{t_2}) +
    eV(t_1) - eV(t_2) \right] \cG  = 0,
\label{UsadelDiff}
\end{equation}
where $D$ is the diffusion coefficient in the cavity and
the Keldysh Green's function satisfies the normalization condition:
\beq
 \sum_l \int_{-\infty}^{+\infty} \!\!\!\!\!\! dt \ \cG_{il}(t_1, t)
\cG_{lj}(t, t_2) = \delta_{ij}\delta(t_1-t_2)
\,.
\label{norm}
\eeq
Since the conductance is controlled by the contacts,
$\cG(t_1,t_2)$ is uniform in the cavity.
\refE{UsadelDiff} can then be integrated on the volume of the cavity,
and relate $\cG$ to the current through the interfaces \cite{nazarov:1999a}.
The equation for $\cG$ takes then the form of a charge
conservation equation:
\beq
    \sum _k { \check {\cal I}_k}
    - {i \tau_D} \left[ i \hbar(\partial_{t_1} + \partial_{t_2}) +
    eV(t_1) - eV(t_2) \right] { \check {\cal G}} = 0
    ,
    \label{usadel}
\eeq
where the voltage in the cavity is given by the electroneutrality condition
\refe{voltage}
and
\beq
    {\check {\cal I}_k} =  {G_Q\over G_\Sigma}
    \sum_n {T_n^{[k]} [  \check  {\cal G}_k, \check  {\cal G}]
    \over 4 + T_n^{[k]}( \{   \check {\cal G}_k, \check {\cal G}\}-2)}
    \label{BC}
\eeq
are the spectral currents that relate the Green's function of the
electrodes $\cG_k$ to the Green function of the cavity \cite{nazarov:1999b}.
The form of $\cG_k$ is given by the equilibrium solution for
metallic leads of \refE{usadel}, explicitly, in presence of a time
dependent potential, one has:
\beq
    \cG_{ok}(t_1,t_2) =
    \left( \begin{array}{cc}
            \delta(t_1-t_2) & 2 F_k(t_1,t_2) \\ 0 & -\delta(t_1-t_2)
            \end{array}
    \right)
    \label{Gnorm}
\eeq
where $\hbar \partial_t \phi_k(t) = eV_k(t)$,
$F_k(t_1,t_2)=e^{i \phi_k(t_1)} F_{eq}(t_1-t_2) e^{-i \phi_k(t_2)}$,
and
$F_{eq}(t) = \int (d\varepsilon/2\pi) e^{i \varepsilon
t/\hbar}\tanh(\varepsilon/2T)]=-i{\cal P} T/\sinh(\pi T t/\hbar)$, with $T$
being the temperature ($k_B=1)$.

In order to calculate the noise we need to modify the Green's
function of the electrodes by introducing two counting fields
\cite{levitov:1996,nazarov:1999a}:
\beq
 \cG_k = e^{i \sigma_x \chi_k/2} \cG_{ok} e^{-i\sigma_x \chi_k/2}
\label{Gchi}
\eeq
where $\sigma_x$ is a Pauli matrix.
Solving \refE{usadel} together with \refE{voltage} gives the value
of $\cG$ from which, using \refE{BC} one can derive the
counting field dependent current:
$
I_k(t,\chi_1,\chi_2) = G_k {\rm Tr}\{\sigma_x \check {\cal I}_k(t,t) \}/2$.
Derivatives with respect to $\chi_1-\chi_2$ give all the moments of
current fluctuations.
We assume that the potential difference between the two leads is
harmonically oscillating at frequency $\Omega$.
Then the time average over the period $2\pi/\Omega$ of
$I_k(t)$ and $-i \partial I_k(t)/\partial (\chi_1-\chi_2)$
gives the DC current and the low frequency noise, respectively.

Let us first discuss briefly the current.
For vanishing $\chi_k$, $\cG$ has the form given by \refE{Gnorm} and
it depends on a single function: $F(t_1,t_2)$.
It is convenient to change gauge by defining
$F(t_1,t_2) = e^{i\phi(t_1)}\,\tilde F(t_1,t_2)\, e^{-i\phi(t_2)}$
where $\hbar \partial_t \phi(t)= e V(t)$, and
introducing new time variables
$t_-=t_2-t_1$ and $t_+=(t_1+t_2)/2$.
$\tilde F(t_+,t_-)$ is a periodic function of $t_+$:
$\tilde F(t_+,t_-) = \sum_n e^{i n \Omega t_+} F_n(t_-)$ and its explicit
expression in terms of the Fourier component of $\tilde F^{(k)}$ comes from
the solution of the equation of motion \refe{usadel}:
\beq
    F_n(t_-) =
    { \sum_k \alpha_k F_n^{(k)}(t_-) \over 1 +  i n \Omega \tau_D}
    \,.
    \label{Fsol}
\eeq
Here
$
F_n^{(k)}(t)= F_{eq}(t) J_n\left(2 A_k \sin (\Omega t/2) \right)
$, $J_n$ are Bessel functions, $A_k=e(V^o_k-V^o)/(\hbar\Omega)$,
and $V^o$ are the amplitudes of the AC field: $V_k(t)=V_k^o \sin(\Omega t)$.
In order to make a connection with the classical description
given by \refE{diffusion} we relate $\tilde F$ to the charge on the
cavity:
\beq
    Q(t_+) = -\left(e/\delta\right) \tilde F(t_- \rightarrow 0,t_+)
    \label{chargedot}
    \,.
\eeq
Usadel equations \refe{usadel} for $t_-\rightarrow 0$ reduces then
to the continuity equation \refe{diffusion} for the total charge.
Note that when the voltage time dependence \refe{voltage}
is enforced, $Q$ vanishes identically, as it can be verified by
calculating the limit $t_-\rightarrow 0$ in \refE{Fsol}.
On the other hand, the energy distribution function,
$\tilde F(t_+,\varepsilon) = \int dt_- \tilde F(t_-,t_+) e^{i \varepsilon t_-/\hbar}$,
varies periodically in time, and its dependence on the AC driving
frequency, $\Omega$, is on the scale of the inverse diffusion time
$\tau_D^{-1}$.
When an electron enters the cavity, after a very short time $\sim \tau_{RC}$, the
charge rearranges to keep the cavity neutral, the distribution function, instead,
will relax on a much longer time given by $\tau_D$.
Since zero-frequency noise probes the electronic distribution function, it should depend
on the frequency $\Omega$ on the same scale.

\begin{figure}
\begin{center}
\includegraphics[width=3.0in]{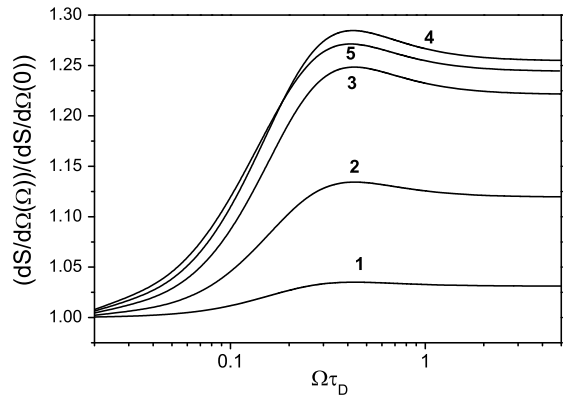}
\caption{ Frequency dependence of a differential photon-assisted shot noise in the
symmetric chaotic cavity ($G_1=G_2$) under fixed
flux $A=e(V^0_1-V^o_2)/(\hbar\Omega)$.
The magnitude of $dS/d\Omega$ is normalized to its value at $\Omega = 0$.
Curve (1) $A = 1.0$, (2) $A = 2.0$, (3) $A = 3.0$, (4) $A = 4.0$, (5) $A = 5.0$.
For symmetric cavity the curves appears to be independent of the
transmission distribution of the contacts.
}
\label{fig2}
\end{center}
\end{figure}

To obtain the noise, it is sufficient to expand the Green's function
up to first order in $\chi$:
$\cG = \cG_o + \cG_1 i (\chi_1-\chi_2)/2+\dots$,
%
%
and substitute this expression into \refE{usadel}
\cite{belzigWire:2001,houzet:2004}.
The term $\cG_o$ is given by \refE{Gnorm} with $F$ given by
\refE{Fsol}.
Note that in principle also $V(\omega)$ could depend on $\chi_k$.
By going back to the Keldysh action formulation we verified that the
corrections are of order higher than one in $\chi_k$, they thus do not contribute
to the calculation of the noise.
In order to fulfil the normalization condition \refe{norm} we use
the parametrization for $\cG_1$ proposed in Ref.~\cite{kamenev:1999}:
\beq
i{\chi}\, \check {\cal G}_1
=
\left( \begin{array}{cc} 1 &F\\
0 & -1 \end{array}\right)
\left( \begin{array}{cc} 0 &
{W} \\ {W'} & 0 \end{array}\right) \left(
\begin{array}{cc} 1 & F \\ 0 &-1 \end{array} \right).
\eeq
By defining $\chi=\chi_1-\chi_2$ and choosing $\chi_{1} = \alpha_2 \chi$,
$\chi_2 = -\alpha_1 \chi$ we obtain  $W'=0$ and the
kinetic equation for $W$:
\beqa
i(1+\tau_D \partial_{t_+})W &=& \sum_k \chi_k \alpha_k
\left[ (1-\beta_k)(F\circ F+F_k\circ F_k)
\nonumber \right. \\
&+&
\left. \beta_k(F\circ F_k+ F_k\circ F) \right],
\label{Wequ}
\eeqa
where $\beta_k=G_Q/G_k \sum_n T_n^{(k)}(1-T_n^{(k)})$ are the Fano
factors of the two junctions and with $\circ$ we indicate time convolution.
To obtain the low frequency noise we need $W$ averaged over one period,
the $n=0$ Fourier component of \refE{Wequ} thus suffices for our purpose.
Substituting the expression for $W$ into the current definition and evaluating the traces
we obtain:
\begin{eqnarray}
\lefteqn{
    S
    =
    G_\Sigma \sum_{n,l,r} \frac{ \hbar\omega_0 (l+r)  \coth \left( {\bar \omega_0 (l+r) \over  2T}\right)}{1+\omega_0^2 \tau_D^2 n^2}
    \times
    }
    \nonumber\\
    &&
    \left[{\cal F}\,{\cal J}(A_1,A_2)+{\cal F}_1\,{\cal J}(A_1,A_1)+{\cal F}_2\,{\cal J}(A_2,A_2)\right]
    \nonumber\\
    &+&
    2 G_\Sigma T (1- \beta_1 \alpha_2 - \beta_2 \alpha_1),
 \label{finalS}
\end{eqnarray}
where
${\cal J}(A_1,A_2)=J_{n+l} (A_1) J_{l} (A_1) J_{r-n} (-A_2) J_{r} (-A_2)$,
${\cal F}_1 = (\alpha_1 + \beta_1 \alpha_2 - {\cal F})/2$,
${\cal F}_2 = (\alpha_2 + \beta_2 \alpha_1 - {\cal F})/2$,
${\cal F}=\alpha_1\alpha_2+\beta_1 \alpha_2^3+\beta_2 \alpha_1^3$,
and the amplitudes are $A_1=\alpha_1 A$, and $A_2=-\alpha_2 A$, with
$ A =(V_1^o-V_2^o)/\hbar \Omega$.
Expression \refe{finalS} is one of the main results of this paper.
For $\Omega \tau_D \ll 1$ it reduces to Lesovik-Levitov result \cite{lesovik:1994}
with the effective Fano factor, ${\cal F}$, appearing at the place of the quantum point
contact Fano factor.
For small $A$ we recover the result of Ref.~\cite{polianski:2005}:
at $A^2$ order there is no dispersion.
For all other cases a frequency dispersion is present, as it can be seen from Fig. \ref{fig2},
where $dS/d\Omega$ is shown as a function of $\Omega$ for small temperature
$T \ll \hbar \Omega$.
In particular, we find that $dS/d\Omega$ displays a weak maximum for $\Omega\sim 1/\tau_D$
reminiscent of the reentrant behavior in superconductors.

%
%
%
%

The driving frequency dependence of the photon-assisted noise may also be generated by
electron diffusion in disordered metals.
To illustrate this point we consider a one-dimensional diffusive wire
of length $L$ and dimensionless conductance $g=G/G_Q \gg 1$.
We employ the same procedure used for a chaotic cavity, with the important
difference that we have to take into account the spatial dependence of $\cG$
in the wire described by \refE{UsadelDiff}.
The condition \refe{voltage} is substituted by the neutrality condition
on the potential along the wire: $eV(z,t) = eV^o \sin(\Omega t)(L-z)/L$,
with $0<z<L$.
We assume now that contact resistance is negligible, thus \refE{UsadelDiff}
is complemented by continuous boundary conditions for $\cG$ at the ends of the wire.
(The electronic energy distribution function for this problem has been
discussed in Ref.~\cite{shytov:2003}.)

To accomplish this program technically we switch to the energy
representation of \refE{UsadelDiff}.
Since the driving is periodic we can single out the $t_+$ dependence for any
operator $\check A$:
$\check A(t_1, t_2) = \sum_n \check A_n(t_{-}) e^{i n \Omega_0 t_+}$.
In the energy domain this implies that
$\check A(\epsilon_1, \epsilon_2) = \sum_n \widetilde A_n(\epsilon_1) \, 2\pi\delta(\epsilon_1-\epsilon_2+ n \hbar\Omega)$,
where $\widetilde A_n(\epsilon - n\hbar \Omega/2 )$ is the Fourier transform of  $\check A_n(t_{-})$.
It is thus convenient to define $\epsilon = E + k \hbar \Omega$, such that $k$ in an integer,
and $-\hbar \Omega/2< E < \hbar \Omega/2$.
One can then represents $\check A$ in matrix form: $\check{A}_{n m}(E) = \widetilde{A}_{m-n}(E + n \Omega)$.
In this way the time convolution between two operators becomes a simple matrix product.
In the matrix representation the distribution function entering \refE{Gnorm} for the right
lead ($z=L$) takes a diagonal form: $F_R(E)_{nm} = \delta_{nm} {\rm sign} (E + n\Omega)$.
For the left lead at zero temperature it reads instead:
\[
     F_L(E)_{nm}
     =
     -\delta_{nm} + {2 \,i^{n-m}} \sum_{l=-\infty}^{m}
     \, J_{l}(-A) J_{m-n-l}(A)
\]
where $A=eV^o/\hbar\Omega$.
In practical calculations it is enough to restrict the matrix size to $|n| \leq 3\,A$.

We solve \refE{UsadelDiff} by representing the wire as a chain of $N-1 \gg 1$ chaotic cavities
(where $\cG$ is uniform) with $N$ identical barriers between them \cite{belzigWire:2001,nazarov:2002b}.
This leads to a finite difference version of the Usadel equation:
\begin{equation}
    \left[
    {1 \over 2}\left(\check{\cal G}_{k-1} + \check{\cal G}_{k+1}\right)
    + i { \check\epsilon - \check V^k  \over N(N-1) \epsilon_{\rm Th}}
        ,
    \check{\cal G}_{k}\right]
    = 0,
 \label{FD_Usadel}
\end{equation}
where $\epsilon_{\rm Th} = \hbar D/L^2$ is the Thouless energy of the wire.
The operators $\check\epsilon$ and $\check V^k$ have  matrix representations
$(\check\epsilon)_{n,n} = E + n\Omega$ and
$\check {V^k}_{nm}(E)=i eV^o (N-k)(\delta_{n,m-1} - \delta_{n,m+1})/(2N)$.
Eq. \refe{FD_Usadel} can be solved by iteration \cite{belzigWire:2001}.
Then photon-assisted shot noise can be obtained by evaluating the $\chi$-dependent current
$I_k(\chi) = g N {\rm Tr} \left\{ [\check{\cal G}_{k},\check{\cal G}_{k+1}]\check\tau_K\right\}/8$
at any one of the $N$ barriers.

\begin{figure}
\begin{center}
\includegraphics[width=3.0in]{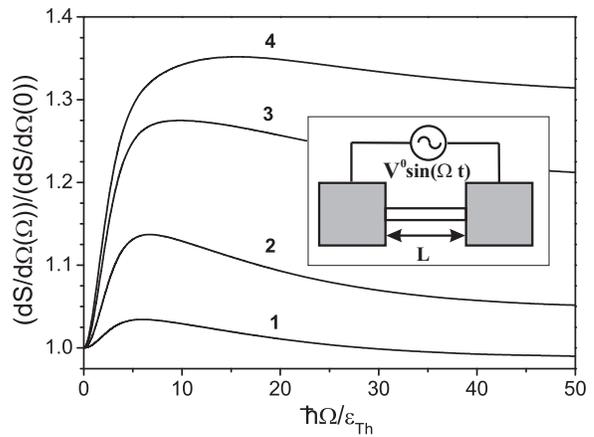}
\caption{
Frequency dependence of the differential photon-noise for a diffusive wire
for given values of $A = eV^o/\hbar\Omega$.
Here $\epsilon_{\rm Th} = \hbar D/L^2$ and the numerical calculation has been performed with
10 nodes.
The magnitude of $dS/d\Omega$ is normalized to its value
at $\Omega = 0$. Curve (1) $A = 1.0$, (2) $A = 2.0$, (3) $A = 3.0$, (4) $A = 4.0$.
}
\label{fig3}
\end{center}
\end{figure}

Results for the differential noise $dS/d\Omega$ versus AC frequency are shown in Fig.~\ref{fig3}.
We find that diffusion due to impurities induces on the photon-assisted noise a similar frequency
dependence as the transmission through a chaotic cavity.
Again a weak maximum is present with the main difference that the energy scale is set by
the Thouless energy instead of the dwell time.
%
%
%

In conclusion, we have shown that the frequency dispersion of the
photon assisted noise can be used to probe directly diffusion times
in mesoscopic conductors.
Our predictions can be verified by an experiment analogous to that
described, for instance, in Ref.~\cite{reydellet:2003},
where a chaotic cavity or a diffusive wire should be substituted
to the quantum point contact.
By carefully choosing the transparencies of the cavity or the length of
the wire, one can match the Thouless energy ($\hbar/\tau_D$) with
the range of frequencies that have been already investigated.
A Thouless energy of 10 $\mu$eV, that is typically realized in mesoscopic
conductors, corresponds to $\Omega/2\pi = 2.4$ GHz which is a
frequency readily accessible in experiments.

We acknowledge useful discussions with F. Hekking, Yu. Nazarov, L. Levitov,
and V. Vinokur.
We acknowledge hospitality of Argonne National Laboratory
where part of this work has been performed
with support of the U.S. Department of
Energy, Office of Science via the contract No. W-31-109-ENG-38.
This work is also a part of research network of
the Landesstiftung Baden-W\"urttemberg gGmbH.


\end{document}